\documentclass[aps,prb,twocolumn,showpacs]{revtex4}

\usepackage{epsfig}
\usepackage{graphics}
\usepackage{subfigure}

\begin{document}

% \markboth{}{CONFIDENTIAL}

%\def\norma#1{\vert \vert {\bf #1} \vert \vert _2}
%\def\cond#1{cond({\bf #1})}
%\def\tcond#1{\widetilde {cond}({\bf #1})}
%\def\dprime#1{ #1^{\prime\prime}}
%\def\abs#1{\vert #1 \vert}
%\def\rerror#1{\norma{\delta #1}}
%\def\ferror#1{\frac{\norma{\delta #1}}{\norma{#1}}}
%\def\derror#1{{\norma{\delta #1}}/{\norma{#1}}}
%\def\TR#1{\parallel \!#1\!\parallel^2}

\def\e{\begin{equation}}
\def\f{\end{equation}}
\def\_#1{{\bf #1}}
\def\o{\omega}
\def\.{\cdot}
\def\x{\times}
\def\E{\epsilon}
\def\va{\varepsilon}
\def\M{\mu}
\def\D{\nabla}
\newcommand{\ds}{\displaystyle}
\def\l#1{\label{eq:#1}}
\def\r#1{(\ref{eq:#1})}
\def\=#1{\overline{\overline #1}}
\def\##1{{\bf#1\mit}}

\title{Giant radiation heat transfer through the micron gaps}

 \author{Igor S. Nefedov, Constantin R. Simovski}
 \affiliation{Aalto University,
School of Electrical Engineering  \\
SMARAD Center of Excellence,
P.O. Box 13000, 00076 Aalto, Finland}

 \date{\today}

 \begin{abstract}

Near-field heat transfer between two closely spaced radiating
media can exceed in orders radiation through the interface of a
single black body. This effect is caused by exponentially decaying
(evanescent) waves which form the photon tunnel between two
transparent boundaries. However, in the mid-infrared range it
holds when the gap between two media is as small as few tens of
nanometers. We propose a new paradigm of the radiation heat
transfer which makes possible the strong photon tunneling for
micron thick gaps. For it the air gap between two media should be
modified, so that evanescent waves are transformed inside it into
propagating ones. This modification is achievable using a
metamaterial so that the direct thermal conductance through the
metamaterial is practically absent and the photovoltaic conversion
of the transferred heat is not altered by the metamaterial.
 \end{abstract}

\maketitle

\section{Introduction}

\begin{figure}[!h]
\subfigure[]{\includegraphics[width=0.35\linewidth]{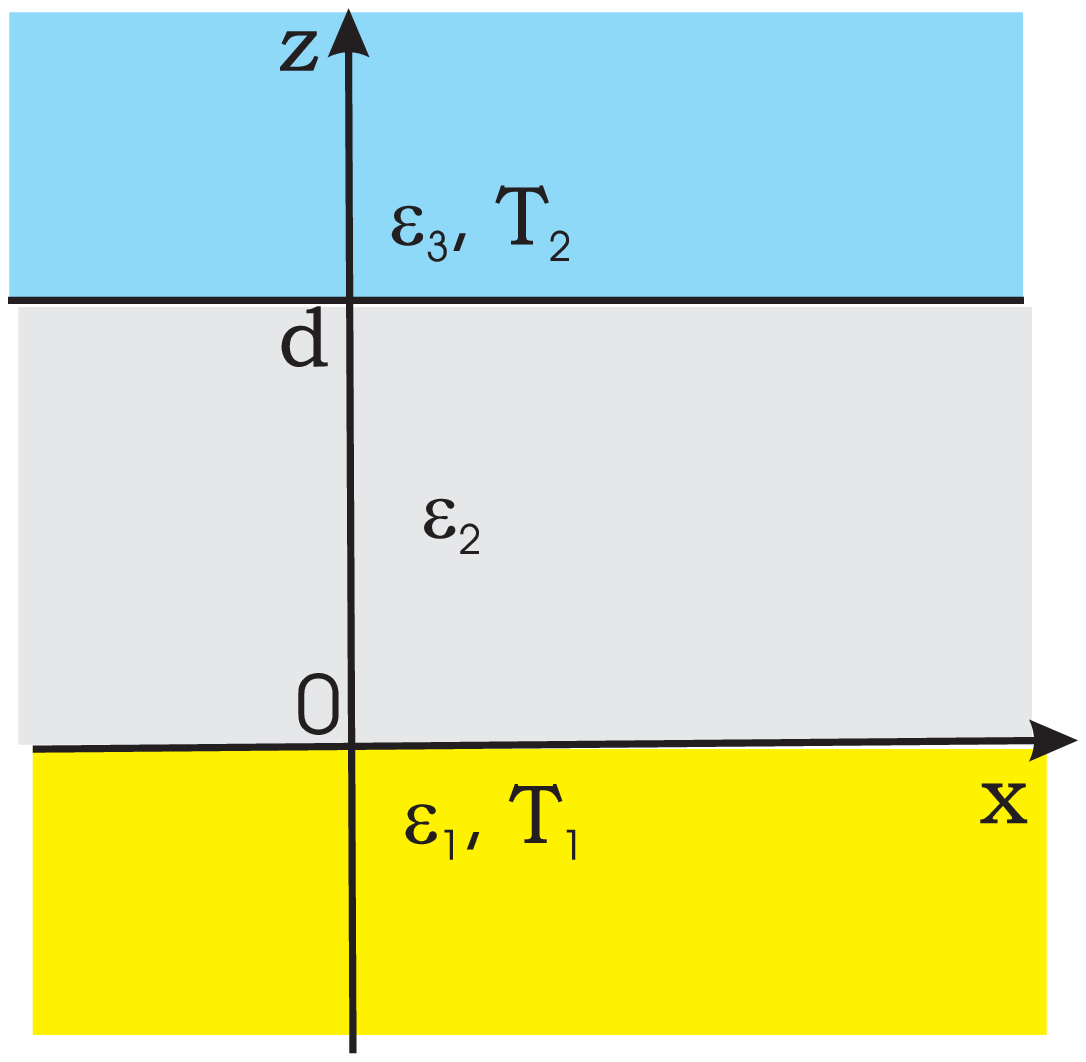}}
\subfigure[]{\includegraphics[width=0.5\linewidth]{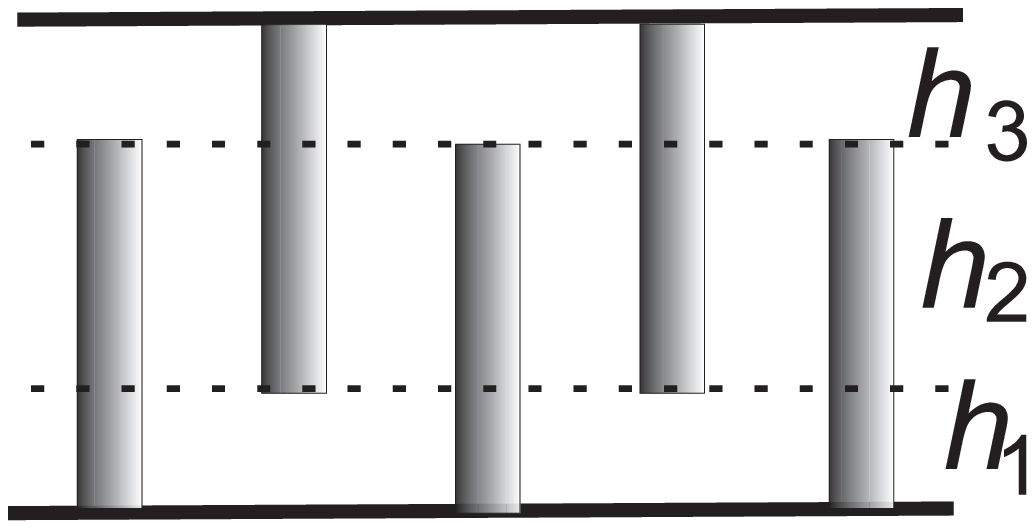}}
%\centering \epsfig{file=schema.eps, width=3cm} \epsfig{file=sxema.eps, width=4cm}
\caption{(Color online) (a) -- Illustration to the general problem
formulation. (b) -- Interdigital arrangement of CNT allows to get
rid of the thermal conductance.}
 \label{sh}
 \end{figure}

Over the last decade near-field thermo-photovoltaic (NF TPV)
systems (e.g. \cite{1,2,3}) are often considered in the modern
literature as a promising tool for the field recuperation from
high-temperature sources (industrial waste heat, car engine and
exhausting pipe, etc.). Also, NF TPV systems are used as precise
temperature profile sensors (e.g. \cite{1,4}). The operation
principle of NF TPV systems is based on the use of the evanescent
spatial spectrum, i.e. on the use of the energy of infrared fields
stored at nanometer distances from the hot surface. In presence of
another body in the near vicinity of the hot surface the
well-known photon tunneling phenomenon arises between two surfaces
which leads to the dramatic increase of the heat transfer compared
to the value restricted by the back-body limit. This transfer can
be enhanced if the hot medium, denoted as medium 1 in Fig.
\ref{sh} (a), and the photovoltaic material, denoted in Fig.
\ref{sh} (a) as medium 3, both possess negative permittivity
\cite{1,5,review,Si}. The enhancement holds at a frequency where
$\rm Re({\va}_{1,3})\approx -1$ (when medium 2 is free space) and
is related to the excitation of coupled surface-plasmon polaritons
(SPP) at the interfaces of media 1 and 3 \cite{5}.

In spite of their relatively high efficiency, NF TPV systems
suffer strong technological drawbacks which restrict their
applicability and industrial adaptation. First, it is difficult to
create flat and clean surfaces of Media 1 and 3 separated with a
nanogap (the roughness should be much smaller than the gap
thickness $d$). Second, the cryogenic cooling or/and vacuum
pumping is required for their operation since the thermal phonons
suppress the photovoltaic conversion \cite{1,review} and the
thermal conduction in the air is very high if $d$ is smaller than
the mean free path of the air molecules ($30-50$ nm) or comparable
with it.

Recently, works on the theory and design of so-called micro-gap
TPV (MTPV) systems have appeared (see e.g. in \cite{MTPV1,MTPV2}).
MTPV systems in which the gap $d$ between the photovoltaic and hot
surfaces is smaller than the operational wavelengthes but
comparable with them occupy an intermediate place between
conventional (far-field) TPV and NF TPV systems. This concerns
also their efficiency which is much less than that of NF TPV
systems though much higher than that of far-field TPV systems. The
lack of efficiency is partially reimbursed by strong technological
advantages. Attempts to increase the efficiency of MTPV systems
are known \cite{1,6,MTPV3} related with the use of nanoantennas
arrays created on both hot and photovoltaic surfaces or insertion
of a photonic crystal layer inside the gap, but the enhancement is
not as significant as to justify these complications. In
principle, one can strongly increase the radiation heat transfer
through the gap partially filling it with a lossless
negative-index metamaterial \cite{new}, however high losses are
inherent to such metamaterials and for substantial gaps and known
metamaterials this mechanism is not very efficient.

In the present paper we suggest a new paradigm for MTPV systems
which should make such structures more competitive than NF TPV
systems. We suggest to fill in the gap between Media 1 and 3 with
the metamaterial 2, performing the transformation of the
evanescent spatial spectrum into propagating one. Metamaterials
performing this manipulation with electromagnetic waves are, e.g.,
the so-called {\itshape indefinite media} \cite{Smith} -- uniaxial
materials in which the axial and tangential components of the
permittivity tensor has different signs. For the infrared range
such metamaterials can be performed as arrays of aligned carbon
nanotubes (CNT) \cite{IgorPRB,arxiv}. Arrays of aligned metallic
CNTs transform incident p-polarized infrared waves (propagating
and evanescent) into quasi-TEM waves propagating along CNTs with
quite small decay. Unlike the SPP enhancement this effect holds
over the whole mid IR range \cite{IgorPRB,arxiv}.

Arrays of single-wall aligned carbon nanotubes with metallic
properties are fabricated by many research groups and used as
field emitters \cite{ShFan}, biosensors \cite{YLin} and antennas
\cite{Wang,DresselNat}. Such parallel CNTs can be few micrometers
long and by two-third of their length free standing (one third is
submerged into a dielectric matrix) \cite{new2}. It is possible to
perform medium 2 so that the direct thermal conduction through
CNTs is avoided. A possible scheme is shown in Fig.~\ref{sh}. The
area filled with medium 2 is divided by three layers:
$d=h_1+h_2+h_3$. Since there is no mechanical contact of any CNT
with both media 1 and 3 and, since the array density is assumed to
be very small, the thermal conductance through the array of CNTs
will be smaller than the thermal conductance through the air gap
(for $d\ge 1\ \mu$m the last one does not lead to the critical
suppression of the photovoltaic conversion).

\section{Theory}

The radiation heat transfer to medium 3 is calculated through the
ensemble-averaged Poynting vector $\langle\_S^{13}_z\rangle$
created by medium 1 at the input of medium 3 (plane $z=d$) minus
the backward Poynting vector $\langle\_S^{31}_z\rangle$
(calculated at $z=0$) \cite{Polder}. The spectral density of the
total heat net transferred between medium 1 with temperature $T_1$
and medium 2 with temperature $T_2$ equals \cite{Polder}: \e
\l{net}
q_{\omega}''=\int_0^{\infty}\left[\langle\_S^{13}_z(k_x,\omega,T_1)\rangle-
\langle\_S^{31}_z(k_x,\omega,T_2)\rangle\right]k_xdk_x. \f In
\r{net} we neglect the heat radiation and absorption in medium 2.
This approximation is justified for CNT arrays of small density
(ratio diameter/period).

To calculate the Poynting vector $\langle\_S^{13}_z\rangle$ (the
value $\langle\_S^{31}_z\rangle$ is obtained similarly) we first
find the incident Poynting vector
$\langle\_S^1_z(k_x,\omega)\rangle$ calculated for the
semi-infinite medium 1 neglecting the reflection from the
interface between media 1 and 2. Then we apply the transmission
matrix method which allows us to express
$\langle\_S^{13}_z\rangle$ through the incident Poynting vector
$\langle\_S^1_z(k_x,\omega)\rangle$ and known parameters of media
2 and 3. Following to \cite{Polder} we start from nonhomogeneous
Maxwell's equations with random electric current density sources
$\_j$ and apply the fluctuation-dissipation theorem (see e.g. in
\cite{1,review}) for the ensemble-averaged bulk current density:
\e\l{fdt}
\begin{array}{lc}
\langle j_m(\_r,\omega)j^*_n(\_r',\omega')\rangle=  \\
 \frac{4}{\pi}\omega\E_0\E''(\omega)\delta_{mn}\delta(\_r-\_r')\delta(\omega-\omega')\Theta(\omega,T).
\end{array}\f
In Eq.~\r{fdt}  $\E''\equiv {\rm Im}(\E_1)$, $j_m$ and $j_n$
($m,\,n=1,\,2,$ or 3) are $x,\ y$ or $z$ component of $\_j$,
$\delta_{mn}$ is the Kronecker symbol, $\delta(x)$ is the Dirac
delta function, and $\Theta(\omega,T)$ is the mean energy of the
Planck oscillator \e\l{po}
\Theta(\omega,T)=\frac{\hbar\omega}{\exp(\hbar\omega/k_BT)-1}\f
where $\hbar$ is the reduced Planck constant, $k_B$ is the
Boltzmann constant, and $T$ is the absolute temperature of the
medium.
%(The term $\frac{1}{2}\hbar\omega$, related to zero field
%vacuum fluctuations is omitted in Eq.~\r{po} since it does not
%contribute to the net radiation heat flux).
According to the ergodic hypothesis, the spectral energy flux of
p-polarized waves is expressed as \cite{ergodic}: \e\l{erg}
\langle\_S^1_z(\_r,\omega)\rangle= \int^{\infty}_0\frac12\langle
{\rm Re}E_x(\_r,\omega) H_y^*(\_r,\omega')\rangle d\omega'
 \f
To find the ensemble-averaged product $\langle E_x H_y^*\rangle$
and integrate it over $\omega'$ is easy taking into account the
homogeneity of the structure in the $(x-y)$ plane. It allows us to
present an elementary bulk current source in a form of a harmonic
current sheet: \e\l{j} \_j(z)=\_j_0(z')\delta(z-z')e^{j(\omega
t-k_xx-k_yy)}. \f Without loss of generality we can put here
$k_y=0$. Solving Maxwell's equations with the source defined by
\r{j} we obtain horizontal field components created at $(x,z)$ by
the source located at $z'$:
 \e\l{ef}\begin{array}{c}
dE_x=\frac{\eta dz'}{2k\E_1}(j_{0x}k_{1z}-j_{0z}k_x)e^{j[\omega t-k_xx-k_{1z}(z-z')]},\\
dH_y=\frac{dz'}{2k_{1z}}(-j_{0x}k_{1z}+j_{0z}k_x)e^{j[\omega
t-k_xx-k_{1z}(z-z')]},\end{array} \f where
$k_{1z}=\sqrt{k^2\E_1-k_x^2}$, $\eta=\sqrt{\va_0/\mu_0}$. The
dependence on $x$ disappears in the cross product \r{erg}. Taking
into account that in \r{fdt} there is no correlation between
different current sheets we can present the integrand in \r{erg}
as: \e\l{ef1}\begin{array}{c} \langle E_x H^*_y\rangle =\frac{\eta
e^{-jk_{1z}z}}{4kk_{1z}\E_1}
\int_{-\infty}^0e^{j(k_{1z}-k_{1z}^*)z'}e^{j(\o-\o')t}\langle
[j_{0x}(\o,z')
\\
k_{1z}-j_{0z}(\o,z')k_x]
[-j_{0x}(\o',z')k_{1z}+j_{0z}(\o',z')k_x]\rangle dz'.\end{array}
\f Further, we can rewrite Eq.~\r{fdt} in the form comprising the
current density amplitudes: \e\l{fdt1}\begin{array}{lc}
\langle j_{0m}(z',\omega)j^*_{0n}(z',\omega')\rangle=  \\
 \frac{4}{\pi}\omega\E_0\E''(\omega)\delta_{mn}\delta(\omega-\omega')\Theta(\omega,T).
\end{array}\f
Substituting \r{ef1} into \r{erg} we easily perform integrating
over $\omega'$ and ensemble averaging using \r{fdt1}. Integrating
over $z'$ we take into account that \e\l{kz}
\int_{-\infty}^0dz'\exp{[j(k_{1z}-k_{1z}^*)z']}=-\frac12\left({\rm
Im}k_{1z}\right)^{-1}.\f The incident Poynting vector created at
the point $z=0$ at the frequency $\omega$ by sources of thermal
radiation distributed in medium 1 yields as follows: \e\l{Sp}
\langle\_S^1_z(k_x,\omega)\rangle= \frac{\E_1''(\omega)/2\pi}{
\E_1 k_{1z}{\rm Im}(k_{1z})
}(k_x^2+k_{1z}k_{1z}^*)\Theta(\omega,T)+{\rm c.c.} \f

Now we have to express the Poynting vector transmitted from medium
1 to medium 3 taking into account the wave reflections from two
interfaces. Assuming that Maxwell's boundary conditions are
satisfied at both boundaries we can present it through the wave
transmission coefficient $\tau$ and transverse wave impedances of
media: \e\l{transm} \langle\_S^{13}_z(k_x,\omega)\rangle=
\frac12\langle\_S^1_z(k_x,\omega)\rangle|\tau|^2\frac{Z_1^*}{Z_3^*}+{\rm
c.c.}, \f where \e\l{imp} Z_i=-E_{ix}/H_{iy}=\eta k_{iz}/k\qquad
(i=1,2,3).\f The coefficient $\tau$ can be obtained through
transmission matrix components of medium 2 in a form \cite{Born}:
\e\l{tr}
 \tau=\frac{2}{M_{11}+M_{12}/Z_3+M_{21}Z_1+M_{22}Z_1/Z_3},\f
where $M_{mn}$ can be found through the z-component $k_{z2}$ of
the wave vector in medium 2: \e\l{comp}\begin{array}{lll}
M_{11}=M_{22}=\cos{k_{2z}d}, & M_{12}=jZ_2\sin{k_{2z}d}, \\
M_{21}=j\frac{1}{Z_2}\sin{k_{2z}d}.
\end{array}\f
If the medium 2 is free space $k_{2z}=-j\sqrt{k_x^2-k^2}$ and for
evanescent waves $|\tau|$ decays exponentially when $k_xd>1$
(beyond the region of SPP \cite{review,Si}). However, medium 2
formed by CNTs even for very large values of $k_x/k$ possess
real-valued $k_z$ (this is so if we neglect losses in CNTs) and
the contribution of spatial frequencies corresponding (in free
space) to the evanescent spectrum is important. In our
calculations we took into account real losses in CNTs, however
these losses practically do not alter the transmission of waves
with large $k_x$.

The applicability of the approach based on the transfer matrix for
finite-thickness slabs of aligned CNTs (practically the validity
of Maxwell's boundary conditions at their boundaries) was checked
in \cite{IgorPRB}. Formula \r{imp} and expression \r{tr} for the
transfer matrix allow high accuracy when calculating the
transmission through a layer filled with a low-density array of
aligned CNTs. Since the backward flux
$\langle\_S^{31}_z(k_x,\omega)\rangle$ calculates similarly to
$\langle\_S^{13}_z\rangle$, all we need to complete our model is
to relate $k_{2z}$ to $k_x$ for given $\o$. For this we used the
model of the indefinite medium with permittivity components
developed for aligned CNTs in \cite{arxiv}. Taking into account
the interdigital arrangement of CNTs depicted in Fig. \ref{sh} the
transfer matrix ${\rm B}$ of the whole gap is calculated as the
product $\rm B=M_1\times M_2\times M_1$, where ${\rm M}_1,\;{\rm
M}_2$ are transfer matrices of layers $h_1$, $h_2$ and $h_3$,
respectively. Matrix $M_2$ corresponds to a twice more dense array
than matrices $M_{1,3}$ do.

%Matrices $M_{1,3}$ are not equal to one another due to the mutual
%horizontal displacement of CNTs.

%Typical dispersion of eigenwaves propagation in arrays of
%single-wall zigzag metallic CNTs forming hexagonal lattice is
%shown in Fig.~\ref{dis}.
%\begin{figure}
% \centering \epsfig{file=HexagonT.eps, width=7cm}
%\caption{(color online). The real part of the slow-wave factor
%${\rm Re}(k_z/k)$, calculated for different lattice periods $a$.}
% \label{dis}
% \end{figure}
%Calculations were implemented numerically using the model
%\cite{IgorPRB} at 20~THz for the array consisting of zigzag
%metallic nanotubes having the radius $r\simeq 0.822$\,nm.
%Figure~\ref{dis} appears similar to a photonic band structure
%diagram, but in contrast the ordinate axis shows the real part of
%the slow-wave factor $k_z/k$. Note, that the slow-wave factor
%equals unity (not zero) at the $\Gamma$ point and the wave remains
%propagating.

Now let us assume that media 1 and 3 have the same permittivity
given by the Drude formula for heavily doped silicon \cite{Si}:
\e\l{Drude}
\E(\omega)=\E_{\infty}-\frac{\omega_p^2}{\omega(\omega-j\gamma)}\f
where $\E_{\infty}\approx 11.6$ is the high-frequency limit value
of the permittivity \cite{Markuier}, and $\gamma$ is the
scattering rate. The plasma frequency and scattering rate are
expressed as $\omega_p=\sqrt{Ne^2/(m^*\E_0)}$ and
$\gamma=e/(m^*\mu$, respectively, where $e$ is the electron
charge, $N$ is the carrier concentration, $m^*$ is the carrier
effective mass, and $\mu$ is the mobility. For $n$-type heavily
doped Si (namely this material is used in NF TPV systems enhanced
by coupled SPP \cite{Si,review}) the mobility expression is given
as \cite{Si} \e\l{mob} \mu=\mu_1+\frac{\mu_{\rm
max}-\mu_1}{1+\left(N_e/C_r\right)^{\alpha}}-\frac{\mu_2}{1+\left(C_s/N_e\right)^{\beta}}.\f
Here $\mu_1=68.5$ cm$^2$/V\,s, $\mu_{\rm max}=1414$ cm$^2$/V\,s,
$\mu_2=56.1$ cm$^2$/V\,s, $C_r=9.2\times 10^{17}$ cm$^{-3}$,
$C_s=3.42\times 10^{20}$ cm$^{-3}$, $\alpha=0711,\;\beta=1.98$,
and $N_e$ is the electron concentration.

\begin{figure}
 \centering
\epsfig{file=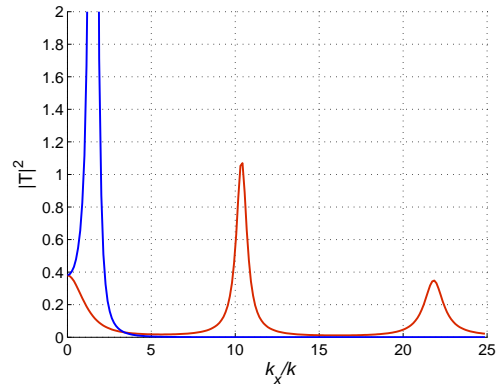,width=0.4\textwidth} \caption{(Color
online) Transmission coefficients versus $k_x/k$ for evanescent
waves in the vacuum gap (blue) and for waves, propagating via the
CNT array (red).}
 \label{Tcomp}
 \end{figure}

Fig.~\ref{Tcomp} presents the comparison of transmittances
 $|\tau|^2$ calculated in presence of the CNT array and in absence
(vacuum gap). Calculations were done at the frequency
corresponding to the wavelength $\lambda=7.5\,\mu$m. Carriers
concentration for $n$-doped silicon was taken
$N_e=10^{20}$\,cm$^{-3}$. The width of the gap between media 1 and
3 was equal in this example $d=500\,nm$ and thicknesses of
sections were as follows: $h_1=h_3=50$\,nm, $h_2=400$\,nm. Periods
of CNT arrays $a_{1,3}=20$\,nm for sections 1 and 3 and
$a_2=10$\,nm for the central section 2. Radius of the CNT was
taken equal $r=0.822$\,nm  and the relaxation time
$\tau_r=10^{-13}$\,s \cite{Achiral}.

%Fig.~\ref{enlarge} shows transmission at small $k_x/k$ in order to
%demonstrate that the points, corresponding to excited plasmons,
%are not singular points at the $k_x$-axis. Peak of transmission
%have finite heights due to losses.
%\begin{figure}
% \centering
%\epsfig{file=enlarge.eps,width=0.4\textwidth}
%\caption{Transmission coefficients at small $k_x/k$ for evanescent
%waves in the vacuum gap (blue) and for waves, propagating via the
%CNT array (red).}
% \label{enlarge}
% \end{figure}

Transmission peaks, exceeding unity, correspond to the excitation
of coupled SPP at the interfaces of media 1 and 3 (compare with
results of \cite{5,6,review,Si}). The most important peak
corresponds to the case of the vacuum gap and is located at
spatial frequencies $k_x/k=(1.5\dots 1.7)$. In presence of CNTs
this peak is suppressed. However, instead, multiple peaks of
$|\tau|$ appear in presence of CNTs and (what is most important)
$|\tau|$, which in absence of CNTs practically vanishes at
$k_x>10k$, in their presence does not. High spatial frequencies
$k_x/k=5\dots 25$ correspond to $|\tau|>0.1$. The limit value
$k_x/k=50$ corresponds to the condition $k_xa_1\simeq\pi/3$. For
spatial frequencies exceeding this limit value the conversion of
evanescent waves into propagating ones disappears (see also in
\cite{IgorPRB}), and corresponding spatial harmonics practically
do not transport energy. Therefore calculating the transferred
heat with formula \r{net} we implement integration over the
Brillouin zone of the CNT lattice with the constant $a_1=20$\,nm.
Obviously, the integral over $k_x$ in formula \r{net} would
diverge if we assume that $k_x\rightarrow\infty$ still correspond
to waves propagating along the CNT axes.

\begin{figure}
 \centering
\subfigure[]{\includegraphics[width=0.45\linewidth]{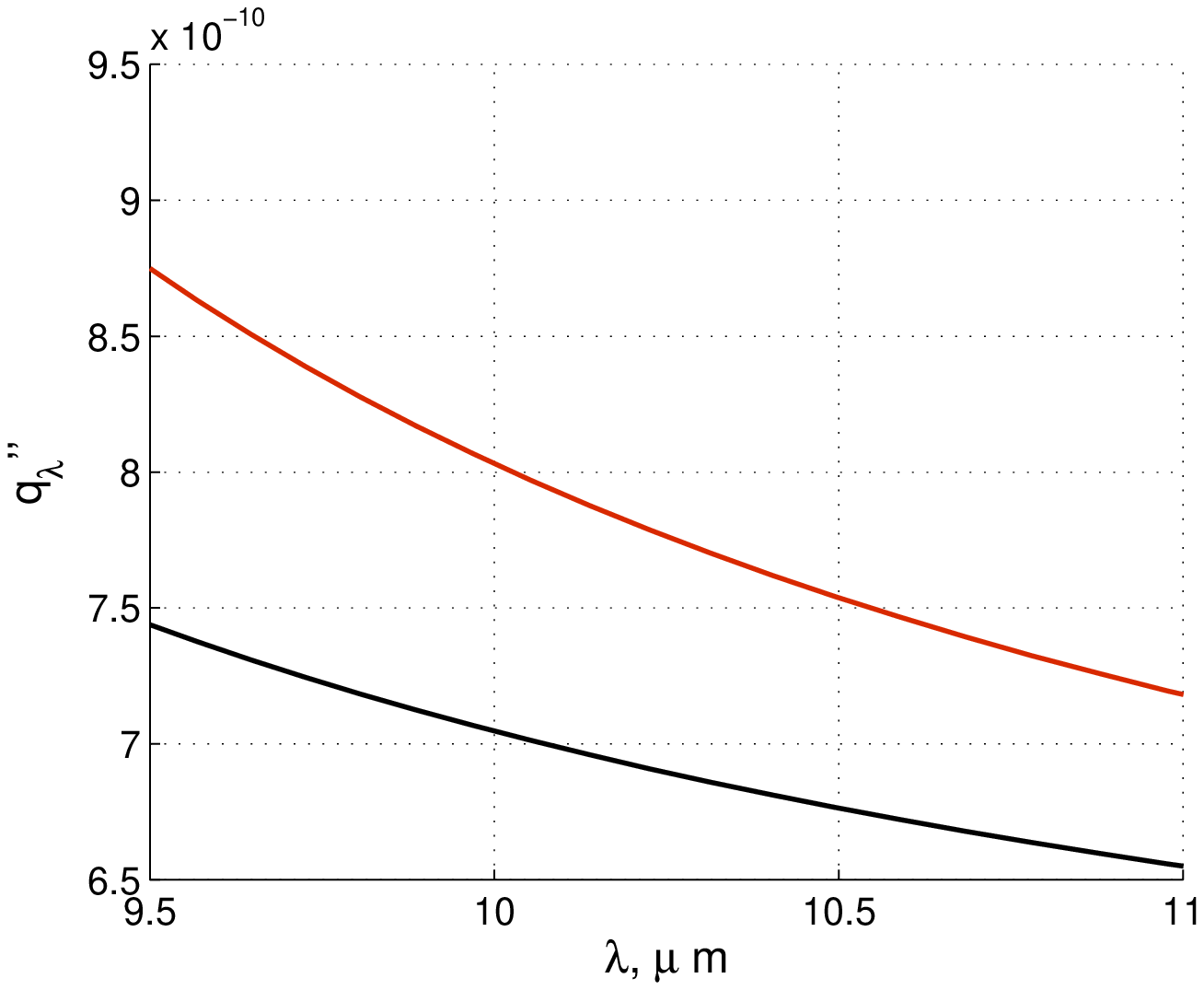}}
\subfigure[]{\includegraphics[width=0.45\linewidth]{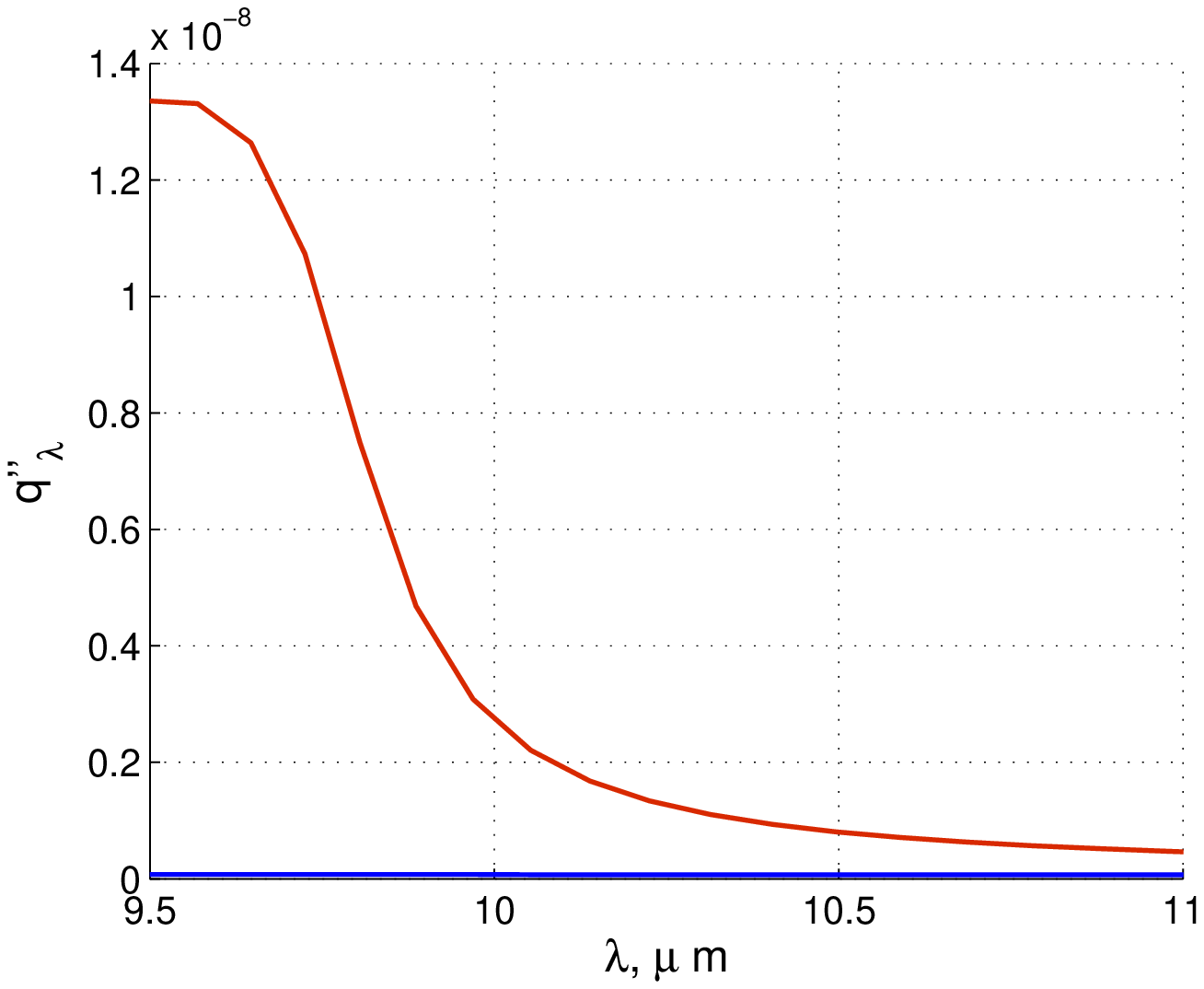}}
\caption{(Color online) $q''$, calculated for the vacuum gap
(blue) and the gap filled with CNTs (red), (a) -- $d=100$ nm, (b)
-- $d=1\ \mu$m.}
 \label{net}
 \end{figure}

Fig.~\ref{net} illustrates the dependence of the spectral density
of the transferred heat $q''_{\lambda}(\lambda)$ (the evident
analogue of $q''_{\omega}$) for the case, when $d=100$\,nm,
$h_1=h_3=10$\,nm and for $d=1000$\,nm, $h_1=h_3=100$\,nm. The
temperatures of the hot and photovoltaic surfaces were in both
examples picked up as follows: $T_1=250^{\circ}$ and $T_2=0$. Then
the maximal heat transfer corresponds to $\lambda_T\approx 9.5$
$\mu$m and $\langle\_S^{31}_z\rangle=0$. For the gap $d=100$ nm
the insertion of CNTs gives a modest overall enhancement about
20\%. This is so because for the gap $d=100$ nm the SPP
enhancement (suppressed by CNTs) is still strong. However, for the
gap $d=1000$ \,nm, where the SPP enhancement is negligible, the
total thermal transfer offered by CNTs exceeds by more than two
orders the heat transfer through the vacuum gap. This effect opens
a new door for the development of the MTPV systems.
\section{Conclusion}
In this study we have shown the way to revolutionary increase the
efficiency of microgap TPV systems inserting between the hot and
the photovoltaic bodies a metamaterial which having the negligible
thermal conductance converts the evanescent waves into propagating
ones in a very broad range of spatial frequencies. We have shown
that this conversion can hold over a wide frequency band and lead
to a dramatic (2-3 orders) enhancement of the total thermal
transfer across the micron gaps.

\end{document}